# Incorrect Data in the Widely Used Inside Airbnb Dataset

Abdulkareem Alsudais
College of Computer Engineering and Sciences, Prince Sattam bin Abdulaziz University, Al-Kharj 11942, Saudi Arabia
E-mail: a.alsudais@psau.edu.sa

**ABSTRACT**

Several recently published papers in Decision Support Systems discussed issues related to data quality in Information Systems research. In this short research note, I build on the work introduced in these papers and document two data quality issues discovered in a large open dataset commonly used in research. Inside Airbnb (IA) collects data from places and reviews as posted by users of Airbnb.com. Visitors can effortlessly download data collected by IA for several locations around the globe. While the dataset is widely used in academic research, no thorough investigation of the dataset and its validity has been conducted. This note examines the dataset and explains an issue of incorrect data added to the dataset. Findings suggest that this issue can be attributed to systemic errors in the data collection process. The results suggest that the use of unverified open datasets can be problematic, although the discoveries presented in this work may not be significant enough to challenge all published research that used the IA dataset. Additionally, findings indicate that the incorrect data happens because of a new feature implemented by Airbnb. Thus, unless changes are made, it is likely that the consequences of this issue will only become more severe. Finally, this note explores why reproducibility is a problem when two different releases of the dataset are compared.

## 1. INTRODUCTION

In a recent paper in this journal, Marsden and Pingry [1, MP] discussed data quality and research reproducibility in Information Systems (IS) and suggested seven questions that should be asked about datasets used in IS research. MP stated that IS research needs to devote additional resources to addressing data quality issues and that published IS papers should detail their data collection process. A subsequent special issue expanded on this subject [2]. The issue included papers that discussed several topics related to data quality in IS [3]–[9]. The primary objective of this short research note is to build on the work presented in MP and the special issue by documenting data quality issues in a popular third-party open dataset commonly used in research. The focus of this note is on the Inside Airbnb (IA) dataset, which is a large open dataset that has been used extensively in research despite not being properly evaluated or thoroughly verified for accuracy. Delving into this matter more deeply, this note challenges the validity of the IA dataset by providing documentation of incorrect data, which may have impacted the findings of peer-reviewed academic papers. These instances of incorrect data increase the number of reviews linked to each listing in IA. In some rare cases, the number of incorrect reviews linked to a listing are higher than the number of actual and correct reviews. The note also explains how MP's question regarding "when" causes reproducibility issues with the IA dataset. These reproducibility issues happen due to IA periodically releasing new versions of listings and reviews. While this is not an issue in the IA dataset itself, it becomes problematic when authors using IA do not specify the release used.

## 2. BACKGROUND ON INSIDE AIRBNB (IA)

IA is a website that was started by an activist who wanted to "dispute Airbnb's claim that 87% of the hosts rent out the place in which they live" [10]. The website offers for direct download data that is reportedly collected from Airbnb's website. Visitors can effortlessly download data from places and reviews from locations such as Los Angeles, New York City, and London. The data for each location includes the date of data compilation. IA collects new data for each location periodically, and a new data for each location replaces all existing ones for the same location. Datasets from IA have enabled scholars to study several relevant topics such as trust in the sharing economy [11]–[13], pricing issues and impacts of the sharing economy [14], [15], and textual contents of reviews extracted from the platform [16], [17]. When searching for the topic "Airbnb" in the Web of Science database and after specifying regular journal articles that were published in 2019, eight of the first 50 papers sorted by relevancy used IA. Additionally, nine other papers used AirDNA, a third-party vendor that collects and sells data scraped from Airbnb. Therefore, 34% of the most relevant papers on Airbnb published in 2019 used an unverified third-party dataset.

While the dataset is widely used in academic research, no independent researchers have provided an evaluation of the dataset and its validity. Additionally, few authors have provided justifications for their use of an open dataset that has not been verified or properly documented. Several authors have stated that they verified the data by selecting a small sample and confirming that the data exist on Airbnb's website [12], [18], [19]. In one of the papers, the sample used for verification contained only 10 entries [18], a sample size that may have been too small for any issues to be discovered. Another justification provided by authors was that the data have been used in other academic papers [15], [16], [20], [21]. The rationale is that the dataset is deemed credible since other authors have used it in their research. However, most of the papers that used IA did not provide justification for their use of an unverified open dataset collected by a third party. Acknowledgements of potential issues in IA's datasets were stated in one paper [22], in which the authors noted possible data quality issues with information collected using a Github project [23] that IA used in their data collection. The authors compared the number of listings collected using the Github project to the number of listings collected by IA. They stated that inconsistencies in the numbers were discovered. This suggests that data quality issues may indeed exist.

Despite the popularity of IA, the website only provides minimal information on the data collection process that was implemented to extract data from Airbnb. According to a page on IA's website [24], the data is collected using python scripts. These scripts included ones that have been "copied and pasted" [24] from other online resources. One such resource is a script available on Github [23]. The person who is responsible for this script indicated that he no longer maintains the project and included a disclaimer that he does not guarantee the quality of the data. Additionally, he documented changes that Airbnb implemented on the layout of their website and explained how these changes negatively affected the web scraper's performance. IA did not provide information on any possible effects these changes had on the performance of their scrapers. However, they noted that some of the reviews in their data might be "spam" added by Airbnb [24]. Websites that are not friendly to scrapers often implement methods to prevent or deceive the scripts [25]. IA stated that these spam reviews were insignificant in terms of size. However, no detailed descriptions or examples were provided on the reviews or the process of their discovery. This lack of clarity regarding the quality of the data raises concerns regarding the use of data from IA.

### 2.1. STRUCTURE OF DATA IN IA

|   | Column | Description |
|---|--------|-------------|
| 1 | Listing_id | The listing ID of the review |
| 2 | ID | The ID of the review |
| 3 | Date | The date of the review |
| 4 | Reviewer_id | The ID of the reviewer |
| 5 | Comment | Textual content of the review |

**Table 1.** Description of Columns in the "reviews" File

Inside Airbnb provides several downloadable files for each available location. The available file downloads for a selected location includes "listings.csv.gz," and "reviews.csv.gz." The "reviews.csv.gz" includes all reviews from the location selected (such as Los Angeles or Rome). Each review is linked to one listing. A listing could, for example, be a room, a house, or an apartment. Several columns that provide more information about the reviews are available. These columns, which are listed and defined in Table 1, include the listing ID, the date the review was written, and the actual textual content of the review. In Airbnb, each listing has a unique listing ID [26] which is used to link to the data in the files "listings" and "reviews." This enables the access of additional information about a listing, such as the name and response rate for the host of the listing, the neighborhood and location (latitude and longitude) of the listing, and the URL of the listing.

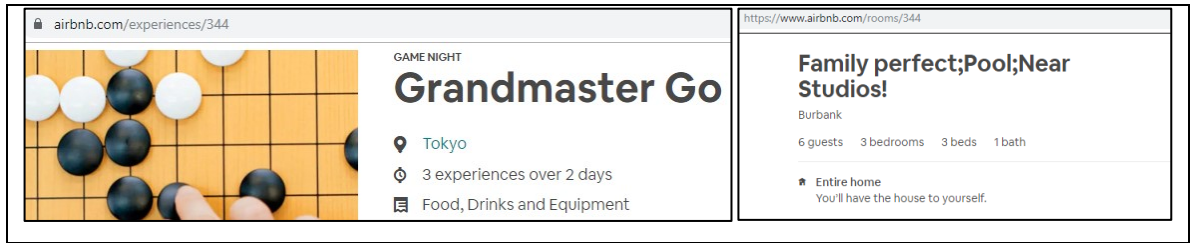

**FIGURE 1.** Two listings with the same listing id "344." First is for an "experience" in Tokyo and the other is for a "place," in this case, a house in Los Angeles.

In November of 2016, Airbnb introduced Airbnb Experiences [27]. This was an attempt by the company to expand its offerings by providing two options for users: 1) places for stays such as rooms, apartments, and houses, and 2) activities and excursions users can book, such as cooking classes, guided city tours, and hiking trips. The structure of URLs for places is http://airbnb.com/rooms/"listing_id_for_place" while it is http://airbnb.com/experiences/"listing_id_for_experience" for experiences. In the December 5th release of the IA dataset for the Los Angeles area, no column exists to indicate that a listing in the table is for a "place" or an "experience." Additionally, based on information from IA's website, there does not seem to be a current option that allows visitors to download information extracted from Airbnb's website about experiences. Upon exploring a representative sample of the data in the downloaded "listings" file, all the listings seem to be of the "place" type. In Airbnb, while it is not possible for two places to have the same listing ID, it is possible to have a "place" and an "experience" with the same listing id. In Figure 1, two listings are displayed. The first is for an "experience" in Tokyo, Japan where users can book a class to learn to play the game "Go," while the second is for a house in Burbank (which is in the Los Angeles, CA, USA area) that is available for rent. Both the house and the class have the same ID, "344." In IA's "reviews.csv.gz" file for the Los Angeles area, all the reviews written by guests who have stayed in the house with the listing ID "344" are available and accurately linked to the listing. However, instances of incorrect data were discovered when it was observed that the reviews for the "Go" class in Tokyo with the listing ID "344" were also added as reviews linked to the house in Los Angeles. This signals a data quality issue that requires additional exploration. The details of this issue are presented in Section 3.1.

## 2.2. DATASET

To investigate data quality issues in data collected by IA, two locations--Los Angeles, CA, USA and Asheville, NC, USA—were selected for examination. The datasets for each location were accessed and downloaded directly from IA. The purpose of the selection of two locations (and thus two sets of files to process for each location) was to determine whether issues discovered were not unique to a single location. According to IA, the Los Angeles dataset was compiled on December 5th, 2019 while the Asheville dataset was compiled on November 28th, 2019. For additional comparison and to study the issue of reproducibility when two different releases of the data from the same location are used, an earlier version of the data from Los Angeles was also tested. The earlier data from Los Angeles will be referred to as Los Angeles 1 (LA1), while the second will be referred to as Los Angeles 2 (LA2). In brief, LA1 and LA2 will be used to assess the issue of reproducibility when two releases of data from the same location are used while LA2 and Asheville will be used to explore the issue of incorrect data and its significance, which is the primary focus of this note. Table 2 shows summary statistics for the three locations. The statistics are based on processing the csv files in "reviews.csv.gz" for each of the three sets. There is an observable decrease in the number of listings and reviews for data from LA1 to LA2. A possible explanation is Airbnb's policy of removing places from the platform if their hosts decide to delete the listings. However, this is only one possible explanation and further investigation is needed for confirmation.

| Location | Los Angeles 1 | Los Angeles 2 | Asheville |
|---|---|---|---|
| Number of unique listings | 35,959 | 32,029 | 2,263 |
| Number of reviews | 1,427,153 | 1,368,997 | 170,973 |
| Max number of reviews per listing | 813 | 813 | 907 |
| Average number of reviews per listing | 39.6 | 42.7 | 75.5 |
| Compilation date (according to IA) | July 8th | Dec 5th | Nov 28th |

**Table 2.** Summary Statistics for the three locations analyzed. Dates for compilation are in 2019.

# 3. ISSUES DISCOVERED

## 3.1. INCORRECT DATA

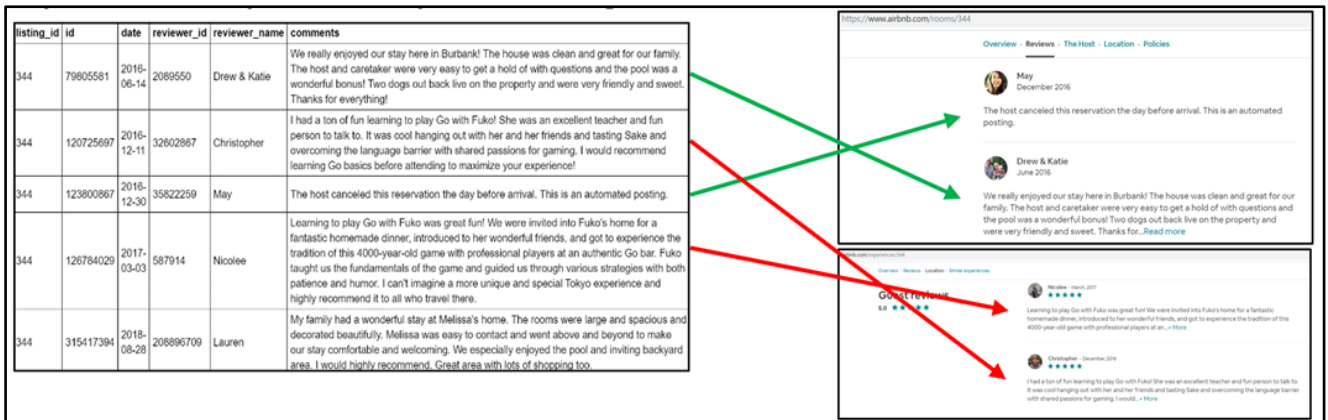

**FIGURE 2.** Example of how reviews in the dataset for the listing ID "343" are actually from two different listings. The ones on top are from the listing of the "place" type with the ID "343" while the ones on the bottom are from the listing for the "experience" with the listing ID "343."

The incorrect data issue pertains to reviews that were incorrectly added to listings in IA. For example, Figure 2 shows five reviews linked to a listing with the listing ID "344." The first and third reviews were both present and found to be accurate when the webpage for the respective listing was visited. However, this was not the case for the second and fourth reviews. While the first review was clearly written by a person documenting their stay in a house in Los Angeles, the second was written by person reviewing a "Go" learning class in Tokyo. As of January of 2020, only two reviews were listed for the experience on the webpage for the class. In the IA dataset for LA2, both reviews appeared as reviews for the house in Los Angeles with the same listing ID (false positive). Moreover, all the reviews for the stay with the same listing ID were also present in the IA dataset (true positive). Therefore, it is possible that an issue in the data collection code written by IA is causing the collection of all the reviews with the specified listing ID regardless of the type of listing ("place" or "experience"). Due to restrictions employed by Airbnb on the automated collection of reviews from places, it is not possible to document in this note if this is indeed the case for all listings. Nevertheless, manual inspections of listings support this hypothesis. Table 3 includes additional examples of incorrect reviews.

To investigate this issue of incorrect reviews found in IA, the files with the reviews from LA2 and Asheville were analyzed. The process, which is illustrated in Figure 3, started by collecting all the listing IDs. Then, for each listing ID, the standard webpage for an Airbnb's experience (http://airbnb.com/experiences/"listing_id_for_experience") was visited. The objective of this step was to see if the ID of the listing was also used as an ID for an experience in Airbnb. This check was completed using a web scraper. The scraper looped over all the available listing IDs and determined whether they are also used as IDs for experiences. To reduce the possibility of issues in the web scraper, a browser window was automatically opened by the script every time a new listing ID was tested in order to monitor the scraper's activity for any errors. One limitation of the scraper is that it is unable to identify if certain IDs were formerly employed as IDs for experiences. Thus, it is possible that the list of IDs that exist as both places and experiences is larger than reported in this note.

| Review | Reported Location | Actual Location |
|---|---|---|
| We really enjoyed our tour with Sara! She knows the history of her country et her city, and she also knows plenty of interesting little stories about art, religion and culture. Alla & Francois | A house in Topanga Canyon (Los Angeles) | Prague, Czech Republic |
| A wonderful excursion with a wonderful local host. Giuseppe was kind, informative, flexible with our schedule and went above and beyond throughout our tour. Highly recommend. | A house in Los Angeles | Agropoli, Italy |

**Table 3.** Examples of incorrect reviews as well as reported locations (by IA) and actual locations.

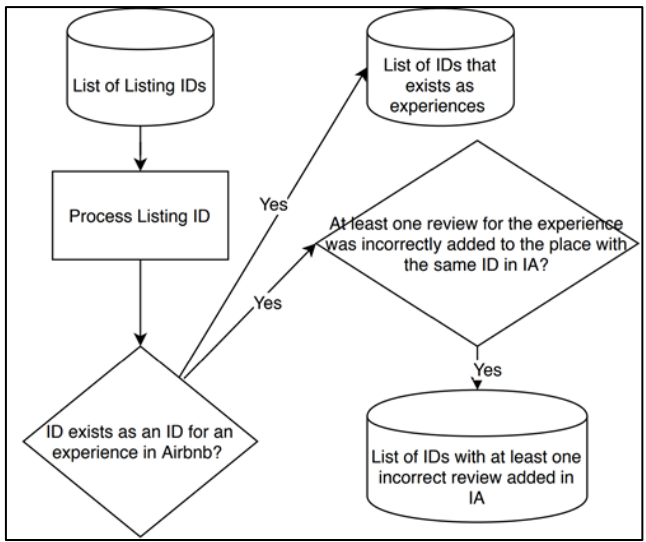

**FIGURE 3.** A flow chart of the process to determine: 1) the IDs that exist as experiences and 2) the IDs in IA with at least one incorrect review.

All the IDs associated with experiences were assessed for accuracy by visiting the URLs of the experiences. All but one ID was found to be accurate. This one ID could be of an experience that has since been deleted. As of January 2020, the "robots.txt" page in Airbnb does not state that they disallow the access of the /experiences/* pages using scripts. For each ID that was classified as one that exists as an experience, the reviews for the experience were examined to determine if at least one review for the experience has been added to the IA dataset. Initially, the objective was to process and compare all the reviews for the place and then the experience with the same ID. However, the "robots.txt" page explicitly states that Airbnb disallows the processing of reviews for a place. This evaluation results in two lists: 1) listing IDs from IA that match both a location and an experience and 2) listing IDs in IA that have at least one review of an experience that was incorrectly added to the listing.

The result of this process is a list of 103 listing IDs that exist as both a listing ID for a place and an experience in LA2, and a list of only five IDs for Asheville. These two lists are available in the appendix. For the 103 listings, at least one review for the experience with the same listing ID as a location was added to the IA dataset for 50 of the IDs. Put differently, 50 of the listings in the IA dataset for LA2 included reviews that were added incorrectly. Some of these reviews were from experiences that include boat rides in India, a yoga class in Australia, and a city tour in Prague. The low number of IDs with incorrect data suggests that while the issue requires attention, it may not currently be severe enough to affect published papers that used the IA dataset. The number of unique listings in LA2 is 32,029, and the percentage of IDs with incorrect reviews is only 0.15%. This low percentage could, however, be due to the overall low number of available experiences. In other words, as more experience and their reviews are added to Airbnb, the quantity of incorrect data in IA is likely to increase. During this inquiry, it was also observed that the mean number of reviews per listing increased from 39.6 reviews per listing in LA1 to 42.7 reviews per listing in LA2.

### 3.2. DIFFERENCES IN RELEASES AND REPRODUCIBILITY ISSUES

While the primary focus of this note is to highlight the issue of incorrect data in the IA dataset, the issue of reproducibility is also of concern. It is imperative that results from one experiment can be reproduced in another. Such replicability is not always possible, however, in studies involving IA that do not specify the version or release used. Since IA releases a new version of the dataset for each location almost every month, the new version replaces an existing release. These updates throw into question the ability for one scholar using the latest release to reproduce results obtained by another researcher who used an older version of the dataset but did not specify the version used. To explore this problem further, two releases for the reviews and listings from the Los Angeles area were analyzed (LA1 and LA2).

The compilation dates as indicated by IA were July 8th, 2019 and December 5th, 2019. By comparing the sizes of the two releases, changes in the number of unique listings and number of reviews were observed. The number of unique listings decreased from 35,959 to 32,029. Thus, 10% of the unique listings in the earlier release were no longer available in the newer release. An attempt to access the webpages of a small random sample of removed listings was made. Results suggest that these listings no longer exist on Airbnb's website, hence it does not seem that there is an issue in the data collection process. However, further exploration is needed for confirmation. For reviews, the number changed from 1,427,153 reviews to 1,368,997 reviews. This suggests that the newer version includes 58,156 fewer reviews. To compare if the change is of statistical significance, Welch's t-test was computed by comparing the number of reviews per listing in the two releases. The null hypothesis in this case is that two releases have an identical average number of reviews per listing. The result of the test was statistically significant ($p$ value < 0.05) and thus the null hypothesis was rejected.

While further examination is needed, this result suggests that significant changes can be observed when comparing the number of reviews in two releases of the dataset for the same location. However, this does not necessarily indicate that these changes are a result of data quality issues in the dataset itself. For this reason, one recommendation for scholars employing the IA dataset in their research is to state the version used in their work in order to avoid potential reproducibility issues. This is also the recommendation of MP who explained the importance of asking the question of "when" a dataset was collected.

## 4. POTENTIAL IMPACT

It is important to assess the impacts that these data quality problems have on published research that used the IA dataset. One challenge to providing such as assessment is the presence of multiple subsets and versions of the dataset. Moreover, based on analyzing a sample of papers that used the dataset, I observed that authors tend to not disclose the versions used. Additionally, for each version used, the entire process detailed in this note to discover incorrect data must be repeated. While this process is largely automated, it still requires manual inspections to confirm the presence of incorrect data due to Airbnb's restrictions regarding automated collection of reviews. Thus, for the incorrect data issue, a process that examines the particular subset used in a paper is required to confirm if findings reported in the paper are still valid and whether incorrect data added had any statistically significant effect.

Still, in order to help gauge the impacts of these findings, all the listing IDs were analyzed to look for any observed patterns. Listing IDs in Airbnb are numerical and in IA the IDs are ordered ascendingly. While no confirmed information was found, it seems that Airbnb creates a new listing ID for new places or experiences by simply adding one to the latest created ID. To further examine the low number of IDs that included incorrect reviews in IA, four classes of IDs were created. The first class included the first 500 IDs in LA2, the second class included the second 500 IDs in LA2, the third class included the third 500 IDs, and the fourth class included the final 500 IDs. Based on matching the IDs that exist as both a place and an experience within these categories, it was found that the first 500 listings included 47 listings, the second 34, the third 22, and the rest of the listings included zero matches. Figure 4 illustrates this decreasing trend. All the results for each sample are in Table 4. No segmentation of the data from Asheville was conducted due to the small number of problematic IDs in the set.

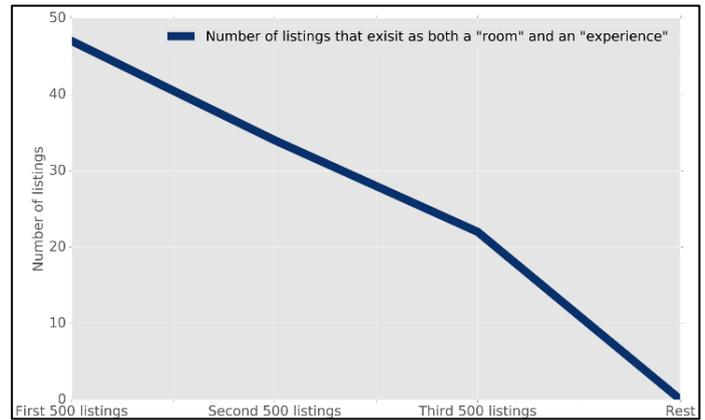

**FIGURE 4.** The number of IDs that exist as both a "room" and "experience" for the four classes of IDs.

While further exploration is needed, these results suggest the following:
- The low number of incorrect data in IA is indirectly related to the overall low number of experiences available in Airbnb. This is supported by the fact that all IDs that exist as experiences are in the first 1,500 IDs of the 32,029 unique listings.
- For the first 500 listings in LA2, incorrect reviews were found for 31 of the listings. That is, 6.2% of the listings included at least one incorrectly added review. This is a significant difference from the 0.15% reported earlier for the entire set. Thus, the effects of this discovered issue could be severe if this subset of the data (the first 500 IDs) is selected.

Therefore, the answer to the crucial question of the statistical significance of the primary discovery depends on how many records of incorrect data were used and for what purpose. For example, if only the first 500 listings were used in a study, the effects could be of statistical significance. Alternatively, if the entire dataset for a set location is used, for example all the reviews from Los Angeles, it is possible that the incorrect data issue will have no effects. It should be noted that an examination of a recent release of the Los Angeles data from IA (compiled on May 8th, 2020) indicates that the incorrect data issue still exists.

| Dataset | Sample | Listings that Exist as Experiences | Listings With at Least One Incorrect Review |
|---|---|---|---|
| LA2 | All | 103 (0.32%) | 50 (0.15%) |
| LA2 | First 500 listings | 47 (9.4%) | 31 (6.2%) |
| LA2 | Second 500 listings | 33 (6.6%) | 13 (2.6%) |
| LA2 | Third 500 listings | 23 (4.6%) | 6 (1.1%) |
| LA2 | Rest (listsings 1501-) | 0 | 0 |
| Asheville | All | 5 (0.22%) | 3 (1.3%) |

**Table 4.** Summary of results for the incorrect data issue

## 5. FINAL OBSERVATIONS

It is worth recalling the work of Marsden and Pingry [1], which proposed seven questions that should be asked before using numerical datasets in IS research. Although IA does not meet the description of the types of numerical data listed in the paper, it would still benefit from being subjected to Marsden and Pingry's seven questions. Therefore, Table 5 provides answers to these seven questions based on 1) what IA provides as answers and 2) what authors generally indicate when they use the dataset. The answers for the "researchers" column are based on examining several papers that used the dataset. Based on these answers, it can be argued that researchers might have avoided using IA had they asked these questions prior to using the dataset in their research. However, further exploration is needed for confirmation. Finally, while no other problems were discovered in IA, it is not guaranteed that the dataset will be free from errors if the issues explained here are addressed. Therefore, it is possible that issues continue to exist in the dataset even after the issues discussed in this note are fixed.

| Questions from [1] | IA | Researchers using IA |
|---|---|---|
| "What provides an explanation of exactly what is captured in the data." | While IA provides an explanation of the dataset, their explanation is limited and lacks detailed descriptions on potential flaws. | Researchers often provide a short description of the dataset but fail to provide a full and detailed explanation. |
| "When refers to the time at which the data is collected." | IA provides the "compilation date" that indicates the date the dataset for a particular location was collected. | Researchers often do not mention the compilation dates for the IA sets used in their work. |
| "Where refers to the location (virtual or real) of the data collection." | IA includes the locations of the data; more specifically, the cities where the listings and reviews originated. | Researchers often do specify the cities used. |
| "How describes the precise process(es) of data collection." | IA does not provide a detailed explanation, but rather shares general information. | Only a few researchers include information on IA's data collection methods or their limitations. |
| "Who details the individual(s) involved in the data collection." | IA shares information about the person responsible for starting the project. | Researchers cite IA but do not include details about individuals involved or their activism and potential biases. |
| "Which details instruments or artifacts used in collecting the data." | IA only shares vague information about the source code used to collect the data. | Researchers often do not describe the methods IA uses to collect the data. |
| "Why provides the set of reasons or goals for collecting the data." | IA explicitly states their primary reason for collecting the data. | Researchers do not indicate that IA started as a project to track the potential negative impacts of Airbnb |

**Table 5.** Answers based on what IA and researchers provide

**REFERENCES**

[1] J. R. Marsden and D. E. Pingry, "Numerical data quality in IS research and the implications for replication," *Decis. Support Syst.*, vol. 115, no. October, pp. A1–A7, 2018.

[2] J. R. Marsden, D. E. Pingry, and J. B. Thatcher, "Perspectives on numerical data quality in IS research," *Decis. Support Syst.*, vol. 126, p. 113172, 2019.

[3] A. Lee-post and R. Pakath, "Numerical, secondary Big Data quality issues, quality threshold establishment, & guidelines for journal policy development," *Decis. Support Syst.*, vol. 126, no. August, p. 113135, 2019.

[4] G. Vial, "Reflections on quality requirements for digital trace data in IS research," *Decis. Support Syst.*, vol. 126, no. August, p. 113133, 2019.

[5] J. Q. Dong, "Numerical data quality in simulation research: A reflection and epistemic implications," *Decis. Support Syst.*, vol. 126, no. August, p. 113134, 2019.

[6] Y. Timmerman and A. Bronselaer, "Measuring data quality in information systems research," *Decis. Support Syst.*, vol. 126, no. August, p. 113138, 2019.

[7] G. Jetley and H. Zhang, "Electronic health records in IS research: Quality issues, essential thresholds and remedial actions," *Decis. Support Syst.*, vol. 126, no. August, p. 113137, 2019.

[8] D. E. O'Leary, "Technology life cycle and data quality: Action and triangulation," *Decis. Support Syst.*, vol. 126, no. August, p. 113139, 2019.

[9] W. Hui, S. M. (Carrie) Lui, and W. K. (John) John, "A reporting guideline for IS survey research," *Decis. Support Syst.*, vol. 126, no. August, p. 113136, 2019.

[10] D. D. Lehr, "An Analysis of the Changing Competitive Landscape in the Hotel Industry Regarding Airbnb," 2015.

[11] L. Zhang, Q. Yan, and L. Zhang, "A computational framework for understanding antecedents of guests' perceived trust towards hosts on Airbnb," *Decis. Support Syst.*, vol. 115, no. October,

pp. 105–116, 2018.

[12] X. Ma, J. T. Hancock, K. L. Mingjie, and M. Naaman, "Self-Disclosure and Perceived Trustworthiness of Airbnb Host Profiles," in *Proceedings of the 2017 ACM Conference on Computer Supported Cooperative Work and Social Computing (CSCW '17)*, 2017, pp. 2397–2409.

[13] L. Zhang, Q. Yan, and L. Zhang, "A text analytics framework for understanding the relationships among host self-description, trust perception and purchase behavior on Airbnb," *Decis. Support Syst.*, vol. 133, no. March, p. 113288, 2020.

[14] J. Chica-olmo, J. G. González-morales, and J. L. Zafra-gómez, "Effects of location on Airbnb apartment pricing in Málaga," *Tour. Manag.*, vol. 77, no. July 2019, p. 103981, 2020.

[15] N. Gurran and P. Phibbs, "When Tourists Move In: How Should Urban Planners Respond to Airbnb?," *J. Am. Plan. Assoc.*, vol. 83, no. 1, pp. 80–92, 2017.

[16] A. Alsudais, "Quantifying the Offline Interactions between Hosts and Guests of Airbnb," in *Twenty-third Americas Conference on Information Systems (AMCIS)*, 2017.

[17] M. Cheng and X. Jin, "What do Airbnb users care about? An analysis of online review comments," *Int. J. Hosp. Manag.*, vol. 76, no. April 2018, pp. 58–70, 2019.

[18] G. Quattrone, A. Greatorex, D. Quercia, L. Capra, and M. Musolesi, "Analyzing and predicting the spatial penetration of Airbnb in U.S. cities," *EPJ Data Sci.*, vol. 7, no. 31, 2018.

[19] A. Alsudais and T. Teubner, "Large-Scale Sentiment Analysis on Airbnb Reviews from 15 Cities," in *Twenty-fifth Americas Conference on Information Systems (AMCIS)*, 2019.

[20] N. Gurran, Y. Zhang, and P. Shrestha, "'Pop-up' tourism or 'invasion'? Airbnb in coastal Australia," *Ann. Tour. Res.*, vol. 81, 2020.

[21] Y. Cai, Y. Zhou, J. (Jenny) Ma, and N. Scott, "Price determinants of airbnb listings: evidence from Hong Kong," *Tour. Anal.*, vol. 24, no. 2, pp. 227–242, 2019.

[22] C. Adamiak, B. Szyda, A. Dubownik, and D. García-Álvarez, "Airbnb Offer in Spain — Spatial Analysis of the Pattern and Determinants of Its Distribution," *ISPRS Int. J. Geo-Information*, vol. 8, no. 3, 2019.

[23] T. Lee, "Airbnb web site scraper," *Github*. [Online]. Available: https://github.com/tomslee/airbnb-data-collection.

[24] "Behind Inside Airbnb." [Online]. Available: http://insideairbnb.com/behind.html. [Accessed: 22-Jan-2020].

[25] R. N. Landers, R. Brusso, K. Cavanaugh, and A. B. Collmus, "A Primer on Theory-Driven Web Scraping: Automatic Extraction of Big Data From the Internet for Use in Psychological Research," *Psychol. Methods*, vol. 21, pp. 475–492, 2016.

[26] M. Haldar, M. Abdool, P. Ramanathan, T. Xu, S. Yang, H. Duan, Q. Zhang, N. Barrow-williams, B. C. Turnbull, B. M. Collins, and T. Legrand, "Applying Deep Learning To Airbnb Search," in *The 25th ACM SIGKDD Conference on Knowledge Discovery and Data Mining (KDD '19)*, 2019, pp. 1927–1935.

[27] Airbnb, "Airbnb Expands Beyond the Home with the Launch of Trips," 2016. [Online]. Available: https://news.airbnb.com/airbnb-expands-beyond-the-home-with-the-launch-of-trips/. [Accessed: 20-Jan-2020].